# Design Method for Electromagnetic Cloak with Arbitrary Shapes Based on Laplace's Equation

**Jin Hu, Xiaoming Zhou and Gengkai Hu[*]**

*School of Science, Beijing Institute of Technology, Beijing 100081, P. R. China*
[*]*Corresponding author: hugeng@bit.edu.cn*

**Abstract:** In transformation optics, the space transformation is viewed as the deformation of a material. The permittivity and permeability tensors in the transformed space are found to correlate with the deformation field of the material. By solving the Laplace's equation, which describes how the material will deform during a transformation, we can design electromagnetic cloaks with arbitrary shapes if the boundary conditions of the cloak are considered. As examples, the material parameters of the spherical and elliptical cylindrical cloaks are derived based on the analytical solutions of the Laplace's equation. For cloaks with irregular shapes, the material parameters of the transformation medium are determined numerically by solving the Laplace's equation. Full-wave simulations based on the Maxwell's equations validate the designed cloaks. The proposed method can be easily extended to design other transformation materials for electromagnetic and acoustic wave phenomena.

## 1. Introduction

Coordinate transformation method for designing electromagnetic materials with a prescribed function has received much attention [1-5]. Based on the form invariance of Maxwell's equations under a coordinate transformation, a topological variation in the transformed space is effectively equivalent to the variation of material parameters in the original space. Through a specific transformation, Pendry *et al* [2] firstly design the well-known spherical and cylindrical cloaks capable of guiding electromagnetic waves around embedded obstacles. The theory is immediately verified by a microwave experiment [4] with help of the metamaterial technology [6], thus opening a new approach by utilizing materials to control electromagnetic waves at will. The transformation method is a general method whatever the shape and function of the transformed media [2,7,8], however the determination of the corresponding material parameters for a given shape and function is a formidable task, which is still an active topics in this field. For cloaks with regular forms, such as square [9] and elliptical [10-13] shapes, they can be designed without much difficulty by constructing explicitly the transformation tensor. To design cloaks with irregular shapes, there are different techniques recently proposed based on the transformation theory [14-17]. Yan *et al.* [14] theoretically show that an arbitrary cloak is possible by analyzing wave property at cloak boundaries. Jiang *et al.* [15] use non-uniform *B*-spline (NURBS) functions to describe the geometrical boundary of a cloak with an irregular shape. Nicolet *et al.* [16]

employ finite Fourier expansions to characterize a random boundary. After a linear mapping, the material parameters of an arbitrary cloak can be obtained. In a different way, Ma *et al*. [17] propose a numerical method by using position surfaces and tracing lines to represent the spatial compression during a linear transformation along the tracing lines. However, the reported works all rely on semi-analytic and semi-numerical approaches to evaluate the transformation matrix **A** for an arbitrary cloak. So a general and flexible method for designing an arbitrary cloak is still necessary. Especially, the answer to the following question needs to be clarified: is there a governing equation which allows us to evaluate the transformed material parameters in a unified way?

In this paper, we will use the Laplace's equation to determine the transformation matrix in a unified manner. It is shown that the function of the transformation media corresponds in fact to specific boundary conditions. So the transformed material parameters are determined by the solutions of the Laplace's equation under proper boundary conditions whatever the shape of a cloak. The paper will be arranged as follows: The design method will be explained in detail in Sec.2. For regular cloaks, the expressions of the permittivity and permeability tensors can be analytically derived with the proposed method. Two examples are given in Sec.3. For irregular cloaks, the proposed method is easily used to compute their material parameters. The full-wave simulations based on the Maxwell's equations validate the effect of the cloak, which will be discussed in Sec.4. The discussion and conclusion will be given in Sec.5.

## 2. Design method

According to the coordinate transformation method [2,8,18], under a space transformation from a flat space **x** to a distorted space $\mathbf{x}'(\mathbf{x})$, the permittivity $\boldsymbol{\varepsilon}'$ and permeability $\boldsymbol{\mu}'$ in the transformed space are given by [18]

$$\boldsymbol{\varepsilon}' = \mathbf{A}\boldsymbol{\varepsilon}\mathbf{A}^{\mathrm{T}} / \det \mathbf{A}, \tag{1a}$$

$$\boldsymbol{\mu}' = \mathbf{A}\boldsymbol{\mu}\mathbf{A}^{\mathrm{T}} / \det \mathbf{A}, \tag{1b}$$

where $\boldsymbol{\varepsilon}$ and $\boldsymbol{\mu}$ are the permittivity and permeability of the original space, respectively. **A** is the Jacobian transformation tensor with the components $A_{ij} = \partial x'_i / \partial x_j$, which characterizes the geometrical variations between the original space $\Omega$ and the transformed space $\Omega'$. The determination of the matrix **A** is the crucial point for designing transformation mediums.

In the continuum mechanics [19], the tensor **A** is called the deformation gradient tensor for an infinitesimal element $d\mathbf{x}$ deformed to $d\mathbf{x}'$ under the space transformation. The transformation can be decomposed into a pure stretch deformation (described by a positive definite symmetric tensor **V**) and a rigid body rotation (described by a proper orthogonal tensor **R**), so the tensor **A** can be expressed as $\mathbf{A} = \mathbf{VR}$ [19]. Suppose that material parameters in the original space are homogeneous and isotropic, and they are expressed by scalar parameters $\varepsilon_0$ and $\mu_0$. Consider the left Cauchy-Green deformation tensor $\mathbf{B} = \mathbf{V}^2 = \mathbf{AA}^{\mathrm{T}}$, Eq. (1) can be rewritten as

$$\boldsymbol{\varepsilon}' = \varepsilon_0 \mathbf{B}/\det(\mathbf{A}),\tag{2a}$$

$$\boldsymbol{\mu}' = \mu_0 \mathbf{B}/\det(\mathbf{A}).\tag{2b}$$

In the principal system, the tensor **B** can be expressed in the diagonal form

$$\mathbf{B} = \mathrm{diag}[\lambda_1^2, \lambda_2^2, \lambda_3^2],\tag{3}$$

where $\lambda_i$ ($i$=1,2,3) are the eigenvalues of the tensor **V** or the principal stretches for an infinitesimal element. Using $\det(\mathbf{A}) = \lambda_1 \lambda_2 \lambda_3$ and Eq. (3), we can rewrite Eq. (2) as

$$\boldsymbol{\varepsilon}' = \varepsilon_0 \mathrm{diag}[\frac{\lambda_1}{\lambda_2 \lambda_3}, \frac{\lambda_2}{\lambda_3 \lambda_1}, \frac{\lambda_3}{\lambda_1 \lambda_2}],\tag{4a}$$

$$\boldsymbol{\mu}' = \mu_0 \mathrm{diag}[\frac{\lambda_1}{\lambda_2 \lambda_3}, \frac{\lambda_2}{\lambda_3 \lambda_1}, \frac{\lambda_3}{\lambda_1 \lambda_2}].\tag{4b}$$

Equation (4) show that the material parameters of the transformation material can be calculated from the principal stretches of the deformation induced by the space transformation. Note that the transformed material parameter and the deformation tensor have the same principal directions and the rigid body rotation in deformations has no contributions on the material parameters. Thus, the calculation of material parameters is converted to evaluate the deformation field for the flat grid element distorted by the space transformation.

Figure 1 shows the scheme for constructing an arbitrary cloak. Suppose an original region enclosed by the outer boundary represented by $b$, inside of this region, we define a point denoted by $a$. Arbitrary cloak can be constructed by enlarging the point $a$ to the inner boundary $a'$, while keeping the outer boundary of the region fixed ($b=b'$). This condition can be expressed by $U'(a) = a'$ and $U'(b) = b'$, where the operator $U'$ is the new coordinates for given point during the transformation. Now the problem is how to determine the deformation fields $\partial U'/\partial \mathbf{x}$ within the cloak layer enclosed by the inner and outer boundary $a'$ and $b'$ for a specific transformation. The commonly used operator $U'$ for designing a cloak is a linear transformation [2,9-13,15-17]. For example, the radial displacement of a spherical cloak is assumed to be a linear relation $r' = (b'-a')r/b' + a'$. However for an arbitrary cloak, it is very difficult to express analytically the boundaries $a'$ and $b'$, the calculation of deformation field is usually very complicated.

To ensure a cloak without reflections, the internal deformation field of the cloak layer must be continuous. The deformation tensor is calculated by the partial derivative of displacement with respect to the original coordinate, so the displacement fields must be smooth enough. It is known that the Laplace's equations with Dirichlet boundary conditions will always give rise to harmonic solutions [20]. This suggests that the displacement field inside of the cloak layer can be calculated by solving Laplace's equations $\Delta U' = 0$ with the boundary conditions $U'(a) = a'$ and

$U'(b) = b'$. In order to keep from the singular solution of the Laplace's equations, we can use the inverse form of the Laplace's equations as

$$\left(\frac{\partial^2}{\partial x_1'^2} + \frac{\partial^2}{\partial x_2'^2} + \frac{\partial^2}{\partial x_3'^2}\right) U_i = 0, i = 1, 2, 3, \tag{5}$$

where $U_i$ denotes the original coordinates in the original space. The corresponding Dirichlet boundary conditions then become $U(a') = a$ and $U(b') = b$. After solving the Eq.(5) with proper boundary conditions, we can get the deformation field inside of the cloak layer which characterizes the distortion of the element. The transformed material parameters can then be calculated from Eq.(4). This method is flexible whatever the shape of cloaks. In the next two sections, we will explain in detail how to obtain analytical expressions of material parameters for regular cloaks and how to design irregular cloaks by solving Eq. (5) numerically.

### 3. Application to regular cloaks

*3.1 Spherical cloak*

The Laplace's equation in the spherical coordinate system is expressed as

$$\left\{\frac{1}{r'^2}\left[\frac{\partial}{\partial r'}\left(r'^2 \frac{\partial}{\partial r'}\right)\right] + \frac{1}{r' \sin \theta'}\left[\frac{\partial}{\partial \theta'}\left(\sin \theta' \frac{1}{r'} \frac{\partial}{\partial \theta'}\right)\right] + \frac{1}{r'^2 \sin^2 \theta'} \frac{\partial}{\partial \varphi'^2}\right\} U_i = 0, i = 1, 2, 3. \tag{6}$$

For a spherical cloak, we let $\theta' = \theta$ and $\varphi' = \varphi$. Then the radial coordinates satisfie the following equation

$$\frac{d}{dr'}\left(r'^2 \frac{dr}{dr'}\right) = 0. \tag{7}$$

With the boundary condition $r(r' = a') = 0$ and $r(r' = b') = b'$ ($b' = b$), the solution of Eq. (7) is given by

$$r = \frac{b'^2}{b' - a'}\left(1 - \frac{a'}{r'}\right). \tag{8}$$

or equivalently

$$r' = \frac{a'b'^2}{(a' - b')r + b'^2}. \tag{9}$$

where $a'$ and $b'$ are respectively the radius of the inner and outer boundaries of the cloak. Equation (9) indicates a nonlinear transformation, which is different from the linear one $r' = (b' - a')r/b' + a'$ commonly used for a spherical cloak [2]. The principal stretches corresponding to the nonlinear transformation (9) are given by

$$\lambda_r = \frac{dr'}{dr} = a'b'^2(b'-a')r', \tag{10a}$$

$$\lambda_\theta = \lambda_\varphi = \frac{r'}{r} = \frac{(b'-a')r'^2}{(r'-a')b'^2}, \tag{10b}$$

From Eq. (4), the material parameters of the spherical cloak are

$$\varepsilon'_r = \mu'_r = \frac{\lambda_r}{\lambda_\theta \lambda_\varphi} = \frac{a'b'^6(r'-a')^2}{r'^3(b'-a')}, \tag{11a}$$

$$\varepsilon'_\theta = \mu'_\theta = \varepsilon'_\varphi = \mu'_\varphi = \frac{\lambda_\theta}{\lambda_\varphi \lambda_r} = \frac{1}{a'b'^2(b'-a')r'}. \tag{11b}$$

Figure 2 shows the electric field distribution of a spherical cloak with the material parameters (11), illuminated by a plane electromagnetic wave. The results validate the invisibility of the designed cloak.

*3.2 Elliptical cylindrical cloak*

In this section, we will design a 2D elliptical cloak by solving the Laplace's equation (5). The Cartesian coordinates $(x, y, z)$ can be expressed by the elliptic cylindrical coordinates $(\xi, \eta, z)$ as

$$\begin{cases} x = c\xi\eta \\ y = \pm c\sqrt{(\xi^2-1)(1-\eta^2)} \\ z = z \end{cases}, 1 \le \xi < \infty, -1 \le \eta \le 1, -\infty < z < \infty, \tag{12}$$

where $c$ is the focal length of the elliptic cylinder. The scale factors of the elliptical cylindrical coordinate system are [21]

$$h_\xi = \frac{c}{\sqrt{\xi^2-1}}\sqrt{\xi^2-\eta^2}, \tag{13a}$$

$$h_\eta = \frac{c}{\sqrt{1-\eta^2}}\sqrt{\xi^2-\eta^2}, \tag{13b}$$

$$h_z = 1. \tag{13c}$$

With Eqs. (12) and (13), the Laplace's equation in the elliptic cylindrical coordinate system is written as

$$\left(\frac{1}{h_{\xi'}^2}\frac{\partial^2}{\partial \xi'^2} + \frac{1}{h_{\eta'}^2}\frac{\partial^2}{\partial \eta'^2} + \frac{\partial^2}{\partial z'^2}\right)U_i = 0. \tag{14}$$

We can construct an elliptic cloak by setting $\eta' = \eta$ and $z' = z$, then Eq. (14) becomes $\partial^2 \xi / \partial \xi'^2 = 0$. With the boundary conditions $\xi(\xi'=a')=1$ and $\xi(\xi'=b')=b$ ($b'=b$), we can obtain a linear transformation relation $\xi' = (b'-a')(\xi-1)/(b'-1) + a'$, where $a'$ and $b'$ are respectively the coordinates of the inner and outer boundaries of the elliptic cloak. The principal stretches for this linear transformation are given by

$$\lambda_\xi = \frac{h_{\xi'} d\xi'}{h_\xi d\xi},\tag{15a}$$

$$\lambda_\eta = \frac{h_{\eta'}}{h_\eta},\tag{15b}$$

$$\lambda_z = 1.\tag{15c}$$

According to Eq. (4), the material parameters of the elliptic cylindrical cloak are given by

$$\varepsilon'_\xi = \mu'_\xi = \frac{\lambda_\xi}{\lambda_\eta \lambda_z} = \frac{\sqrt{\xi^2-1}}{\sqrt{\xi'^2-1}}\frac{b'-a'}{b'-1},\tag{16a}$$

$$\varepsilon'_\eta = \mu'_\eta = \frac{\lambda_\eta}{\lambda_z \lambda_\xi} = \frac{\sqrt{\xi'^2-1}}{\sqrt{\xi^2-1}}\frac{b'-1}{b'-a'},\tag{16b}$$

$$\varepsilon'_z = \mu'_z = \frac{\lambda_z}{\lambda_\xi \lambda_\eta} = \frac{\xi^2-\eta'^2}{\xi'^2-\eta'^2}\frac{\sqrt{\xi'^2-1}}{\sqrt{\xi^2-1}}\frac{b'-1}{b'-a'}.\tag{16c}$$

where we have assumed $\varepsilon_0 = \mu_0 = 1$. It can be found that Eq. (16) agrees with the results given by Ma *et al* [10] from a different method. We can find that the proposed method can give a unified way for designing cloaks by solving the Laplace's equation with proper boundary condition.

### 4. Application to irregular cloaks

For an arbitrary cloak, its boundary is difficult to be expressed in an analytical form, so the Laplace's equation must be solved numerically. In the following, we will show how to use the proposed method to design a cloak of irregular shapes with help of the commercial software COMSOL Multiphysics. Firstly, we depict two boundaries $a'$ and $b'$ for an arbitrary cloak and set the boundary conditions $U(a') = 0$ and $U(b') = b$ ($b' = b$), then solve the Laplace's equations to obtain the deformation field $\partial U / \partial \mathbf{x}'$ inside of the cloak layer. The material parameters of the cloak can be calculated numerically by the principle stretches $\lambda_1$, $\lambda_2$, and $\lambda_3$ (evaluated from the deformation field) according to Eq. (4). This can be achieved with the PDE solver provided by the commercial software COMSOL Multiphysics. In order to check the performance of the designed cloak, the cloak device will be illuminated by a plane electromagnetic wave with help of the RF module in the same software. So the design and validation of the cloak can be well integrated in a two-step modeling.

As an example, we define a two-dimensional arbitrary cloak, which boundaries are formed by the circles of radii 1cm and 3cm, as shown in Fig.3. The cloak is embedded in the air of a square area surrounded by the PML regions. In the PDE modes of Laplace's equation, we set the boundary conditions $U(a') = 0$ and $U(b') = b$ ($b' = b$) for the inner and outer boundaries, then solve the equation $\Delta' U = 0$ to get the coordinates $\mathbf{x} = U(\mathbf{x}')$ [$\mathbf{x}' \in (a', b')$], from which we obtain $\mathbf{x}' = U'(\mathbf{x})$ [ $\mathbf{x} \in (0,b)$ ]. Figure 4 gives the coordinate lines of $\mathbf{x}'$ transformed from a flat space $\mathbf{x}$. The line path implies that electromagnetic waves will propagate around the area bounded by $a'$ and any objects put inside of this area will be invisible. For the verification, we let a plane electromagnetic wave of frequency 20 GHz with the electric field polarized along the $z$ direction illuminate the cloak. The full fields of the system are solved with help of the TE waves mode of COMSOL Multiphysics. Fig. 5 (a) and 5(b) shows the contour plots of the electric field $E_z$ for the waves incident on the cloak horizontally and at an angle of 45° from the $x$ direction, respectively. It can be seen from the figures that the constructed arbitrary cloak doesn't disturb the incident waves and can shield an irregular obstacle from detection. For the physical realization, one can easily retrieve the material parameters of the arbitrary cloak from the deformation field according to Eq. (4). The values of some components of the permittivity and permeability tensors, $\varepsilon_{zz}$, $\mu_{xx}$, $\mu_{xy}$, and $\mu_{yy}$ of the cloak are shown in Figs. 6(a), 6(b), 6(c), and 6(d), respectively. It can be found that the cloak are highly anisotropic and must be realized with the metamaterial technology.

The proposed method can also be used to design arbitrary cloaks with many separated embedded obstacles. For example, we use the same system as shown in Fig.3, but separate the inner circles of the radius 1cm apart, as shown in Fig. 7. The coordinates of central points of the four circles in centimeters are $O_1(0,1)$, $O_2(1,0)$, $O_3(0,-1)$, and $O_4(-1,0)$. The boundary conditions for numerically solving the Laplace's equation then become $U(a_i') = O_i (i=1,2,3,4)$ and $U(b') = b$. Figures 8 (a) and 8(b) show the contour plots of the electric fields $E_z$ for the waves incident on this cloak horizontally and at an angle of 45° from the $x$ direction, respectively. We can see from the figures that the perfect invisibility is still achieved even if the embedded obstacles are separated apart.

Finally the proposed method is applied to design the cloaks with the boundaries drawn randomly or imported from other tool packages, such as Microsoft Office Visio by DXF files. In the following example, we draw an arbitrary cloak at the user-guide interface in the COMSOL software. After solving the Laplace's equation with the boundary conditions for the cloak, we calculate the distorted coordinate lines, as shown in Fig. 9. Figures 10(a) and 10(b) show the contour plots of electric fields $E_z$ for TE electromagnetic waves impinging the cloak horizontally and vertically respectively. It can be seen that the designed cloak does not disturb the outside fields and shields the inner random object from detection.

## 5. Conclusion

Based on coordinate transformation method, a unified method is proposed to evaluate the transformed material parameter for arbitrary cloaks. In this method, the calculation of the material parameters for a cloak becomes the computation of deformation field governed by the Laplace's equation with proper boundary conditions. The principle stretches induced by the transformation are related to the permittivity and permeability tensors in the principle system. By imposing different boundary conditions, we can design arbitrary cloaks with single or many separated embedded obstacles. For regular cloaks with spherical, cylindrical and elliptical shapes, explicit expressions of the material parameters can be easily obtained by solving the Laplace's equation analytically. The design and full-wave simulation of a cloak can be well integrated in a two-step modeling. It is worth to note that the proposed method is not just limited to the design of cloaks, it can also be extended to design other transformation materials, such as rotators [22], concentrators [9], beam shifter [23] and beam bender [24] by the modification of boundary conditions. In addition, we can use this method to design acoustic transformation materials. These issues are currently under investigation.

## Acknowledgments

Dr. Xiaoning Liu is acknowledged for helpful discussions. This work is supported by the National Natural Science Foundation of China (90605001, 10702006, 10832002), and the National Basic Research Program of China (2006CB601204).

## References

[1] Greenleaf, M. Lassas, and G. Uhlmann, "On non-uniqueness for Calderón's inverse problem," Math. Res.Lett. **10**, 685 (2003).
[2] J. B. Pendry, D. Schurig, and D. R. Smith, "Controlling Electromagnetic Fields," Science **312**, 1780 (2006).
[3] U. Leonhardt, "Optical conformal mapping," Science **312**, 1777 (2006).
[4] D. Schurig, J. J. Mock, B. J. Justice, S. A. Cummer, J. B. Pendry, A. F. Starr, and D. R. Smith, "Metamaterial Electromagnetic Cloak at Microwave Frequencies," Science **314**, 977 (2006).
[5] W. Cai U. K. Chettiar, A. V. Kildishev, and V. M. Shalaev, "Optical cloaking with metamaterials," Nat. Photonics **1**, 224 (2007).
[6] D. R. Smith, J. B. Pendry, and M. C. K. Wiltshire, "Metamaterials and Negative Refractive Index," Science **305**, 788 (2004).
[7] U. Leonhardt, "Notes on conformal invisibility devices," New J. Phys. **8**, 118 (2006).
[8] U. Leonhardt, "General relativity in electrical engineering," New J. Phys. **8**, 247 (2006).
[9] M. Rahm, D. Schurig, D. A. Roberts, S. A.Cummer, and D. R. Smith, "Design of Electromagnetic Cloaks and Concentrators Using Form-Invariant Coordinate Transformations of Maxwell's Equations," Photonics Nanostruc. Fundam. Appl. **6**, 87 (2008).
[10] H. Ma, S. B. Qu, Z. Xu, J. Q. Zhang, B. W. Chen, and J. F. Wang, "Material parameter equation for elliptical cylindrical cloaks," Phys. Rev. A **77**, 013825 (2008).
[11] Y. You, G. W. Kattawar, P. W. Zhai, and P. Yang, "Invisibility cloaks for irregular particles using coordinate transformations," Opt. Express **16**, 6134 (2008).
[12] D. Kwon and D. H. Werner, "Two-dimensional eccentric elliptic electromagnetic cloaks," Appl. Phys. Lett. **92**, 013505 (2008).
[13] W. X. Jiang, T. J. Cui, G. X. Yu, X. Q. Lin, Q. Cheng and J. Y. Chin, "Arbitrarily elliptical–cylindrical invisible cloaking," J. Phys. D: Appl. Phys. **41,** 085504 (2008).
[14] W. Yan, M. Yan, Z. Ruan and M. Qiu, "Coordinate transformations make perfect invisibility cloaks with arbitrary shape," New J. Phys. **10**, 043040 (2006).
[15] W. X. Jiang, J. Y. Chin, Z. Li, Q. Cheng, R. Liu, and T. J. Cui, "Analytical design of conformally invisible cloaks for arbitrarily shaped objects," Phys. Rev. E **77**, 066607 (2008).
[16] Nicolet, F. Zolla, and S. Guenneau, "Electromagnetic analysis of cylindrical cloaks of an arbitrary cross section,"Opt. Lett. **33**, 1584 (2008).
[17] H. Ma, S. Qu, Z. Xu, and J. Wang, "Numerical method for designing approximate cloaks with arbitrary shapes," Phys. Rev. E **78**, 036608(2008).
[18] G. W. Milton, M. Briane and J. R. Willis, "On cloaking for elasticity and physical equations with a transformation invariant form," New J. Phys. **8**, 248(2006).
[19] W. M. Lai, D. Rubin and E. Krempl, *Introduction to Continuum Mechanics*, 3 edition. (Butterworth-Heinemann, Burlington, 1995).
[20] R. Courant, D. Hilbert, *Methods of Mathematical Physics*, Vol.2, 1 edition. (Wiley-Interscience ,New York, 1989).
[21] G. Arfken, *Mathematical Methods for Physicists* (Academic Press, Orlando, 1970).
[22] H. Chen and C. T. Chan, "Transformation media that rotate electromagnetic fields," Appl. Phys. Lett. **90**, 241105 (2007).
[23] M. Rahm, S. A. Cummer, D. Schurig, J. B. Pendry, D. R. Smith, "Optical Design of Reflectionless Complex Media by Finite Embedded Coordinate Transformations," Phys. Rev. Lett. **100**, 063903 (2008)
[24] M. Rahm, D. A. Roberts, J. B. Pendry and D. R. Smith, "Transformation-optical design of adaptive beam bends and beam expanders", Opt. Express **16**, 11555 (2008).

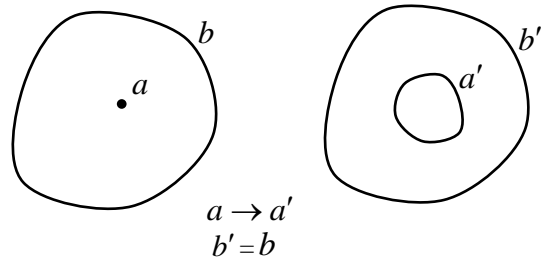

Fig. 1. The scheme of constructing an arbitrary cloak.

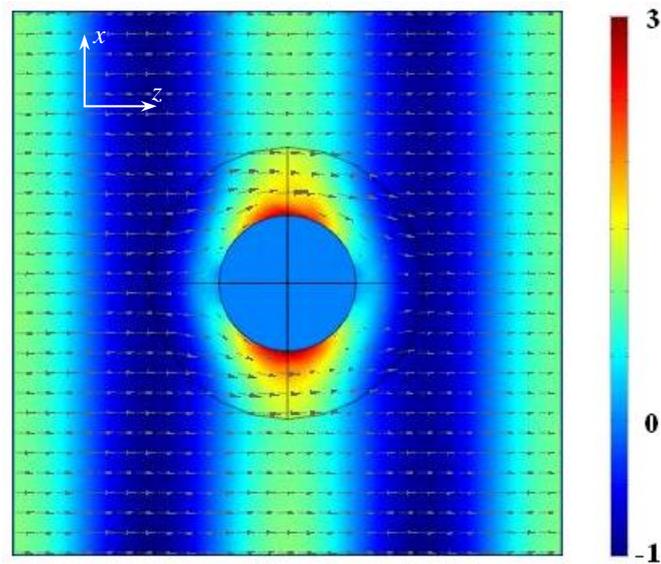

Fig. 2. The contour plot of electric fields $E_x$ in the $x$-$z$ plane for a plane electromagnetic wave of the electric field polarized in the $x$ direction incident on the cloak along the $z$ direction, which material parameters are given by Eq. (11). (The inner and outer radii of the cloak are $\lambda/4$ and $\lambda/2$ respectively, where $\lambda$ is the wavelength, gray arrows indicate directions of the power flow).

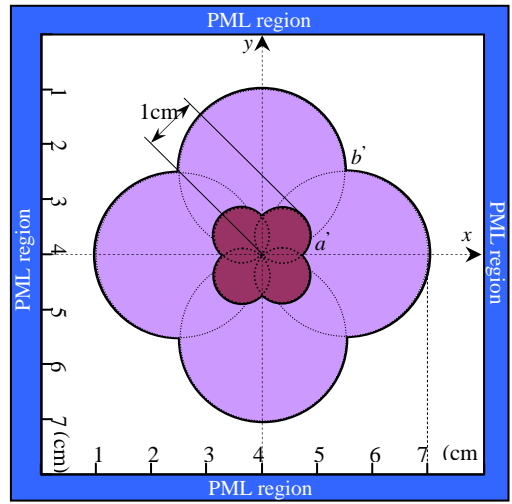

Fig. 3. The simulation environment of a 2D arbitrary cloak with the inner boundary $a'$ and outer boundary $b'$ illuminated by a plane electromagnetic wave of frequency 20GHz.

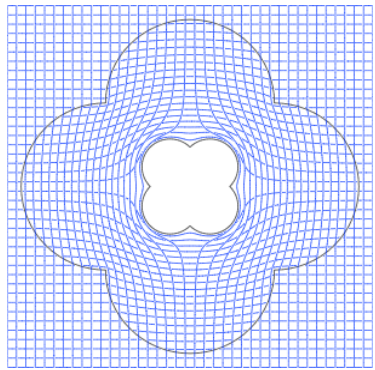

Fig. 4. The coordinate lines in the transformed space distorted by the arbitrary cloak shown in Fig. 3.

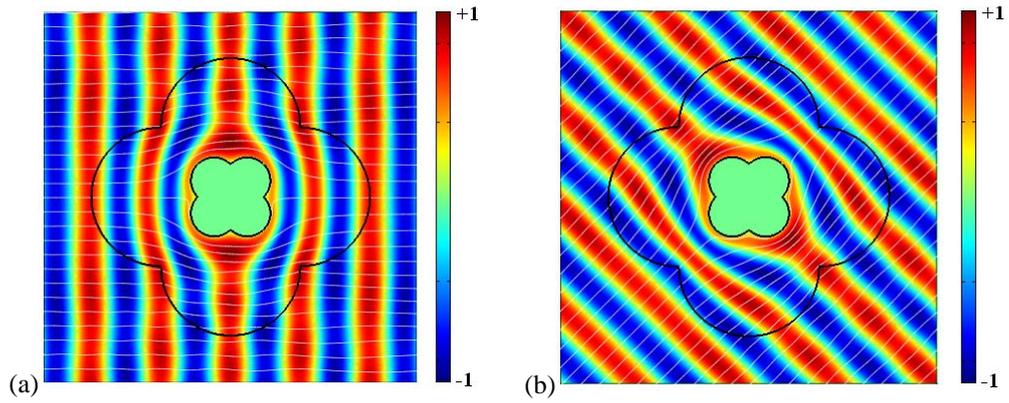

Fig. 5. The contour plots of electric fields $E_z$ for the TE waves incident on the cloak in Fig. 3, (a) horizontally and (b) at an angle of $45°$ from the $x$ direction. (The white lines indicate directions of the power flow).

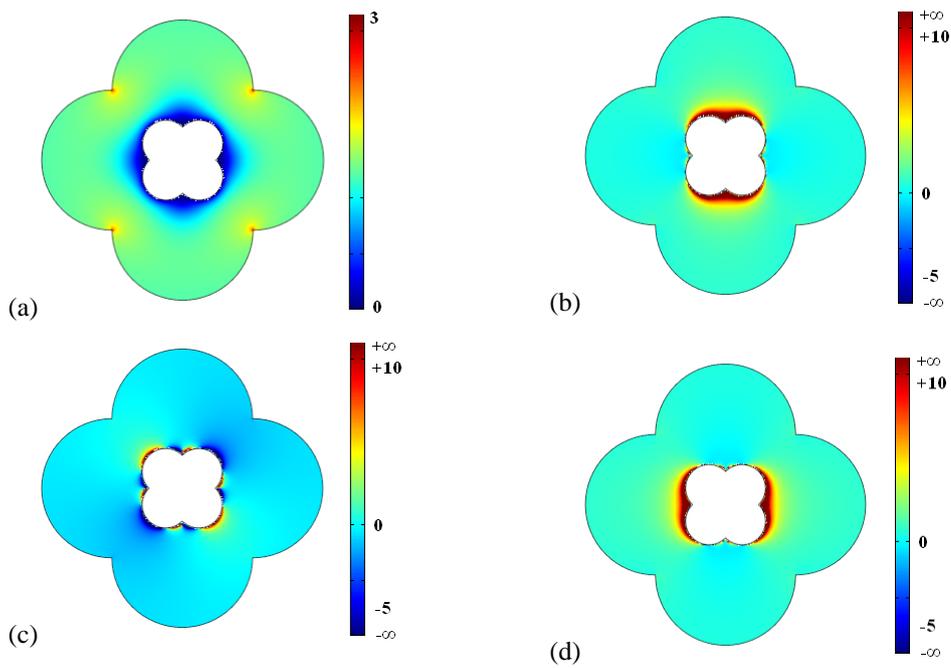

Fig. 6. The contour plots of material parameters (a) $\varepsilon_{zz}$, (b) $\mu_{xx}$, (c) $\mu_{xy}$, and (d) $\mu_{yy}$ of the cloak.

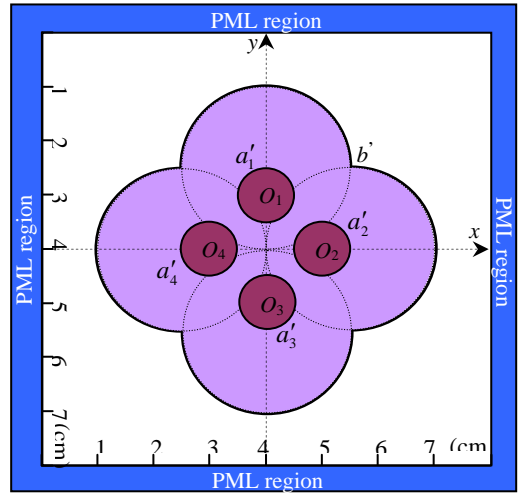

Fig. 7. The scheme of a 2D arbitrary cloak with separated embedded obstacles. (The simulation environment is the same as that used in Fig. 3).

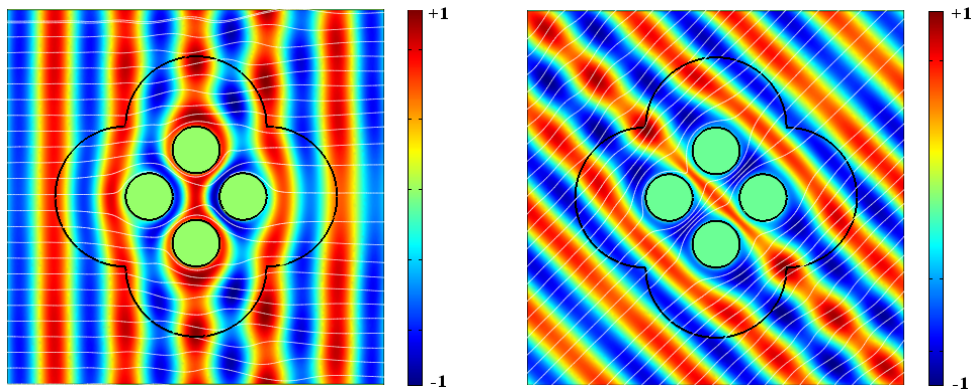

Fig. 8. The contour plots of electric fields $E_z$ for the TE waves incident on the cloak in Fig. 7 horizontally (a) and at an angle of $45°$ from the x direction (b). (The white lines indicate directions of the power flow).

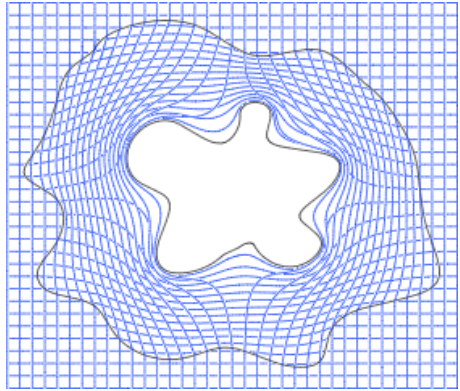

Fig. 9. The coordinate lines in the transformed space distorted by an arbitrary cloak.

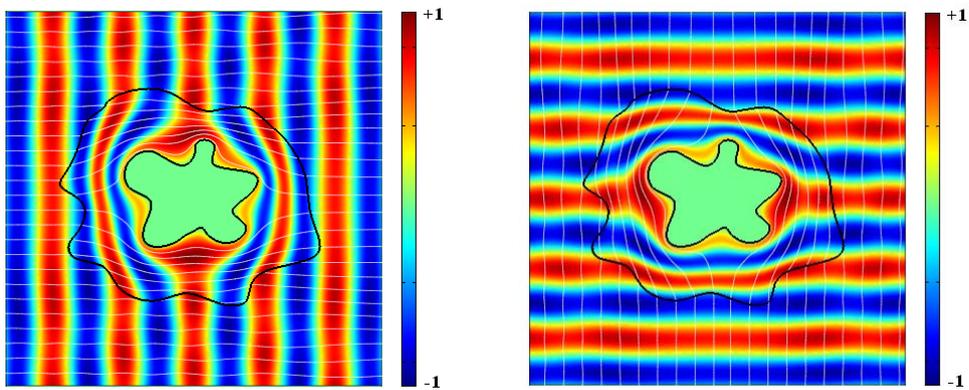

Fig. 10. The contour plots of electric fields $E_z$ for the TE waves of frequency 20 GHz incident on an arbitrary cloak (a) from left to right and (b) from bottom to top. (The size of square simulation region is 8cm×8cm and the white lines indicate directions of the power flow).